\title{Steady energy transfer dependence granular temperature on single bouncing granular particle}
\author{
Suparno Satira\\
Theoretical High Energy Physics Reasearch Division\\
Institut Teknologi Bandung\\
Jalan Ganesha 10, Bandung 40132, Indonesia\\
\and
Sparisoma Viridi\\
Nuclear Physics and Biophysics Research Division\\
Institut Teknologi Bandung\\
Jalan Ganesha 10, Bandung 40132, Indonesia\\
\and
Freddy P. Zen\\
Theoretical High Energy Physics Reasearch Division\\
Institut Teknologi Bandung\\
Jalan Ganesha 10, Bandung 40132, Indonesia\\
}
\date{\today}

\documentclass[12pt]{article}
\usepackage{graphicx}
\usepackage{amssymb}

\begin{document}
\maketitle

\begin{abstract}
Simulation of a system consisted of free particle bouncing on a vertically vibrated based is performed. Two different states, which are steady and unsteady energy transfer state are found. The vibrating based is hold at constant vibration frequency $f = 0.1$ as the vibration amplitude $A$ varied. Sinusoidal form is used. Granular temperature $T_g$ as function of based velocity and coefficient of restitution is used but shown no role in determining energy transfer state of the system. Peak of free particle trajectory $x_m$ around value 20 seperate region of the two states.
\medskip \\
{\bf Keywords:} granular temperature, single particle, steady-unsteady energy transfer.
\end{abstract}

\section{Introduction}
System of a bouncing particle on a vertically vibrated based is not so simple as it sounds and it can be seen as part of more complex system of granular materials. Granular temperature of the system is scaled to based velocity and coefficient of restitution \cite{Warr_1995}. In randomly vibrating based it exibits an inelastic collapse \cite{Majumdar_2007}. Study of stability of existence of periodic modes is also conducted \cite{Barroso_2009}, which triggered further behavious characterization numerically \cite{Macau_2010}. Completely inelastic particle is also subject of similar system \cite{Gilet_2009}. A steady bounce mode and unsteady one are observed in simulation, which is explained by term of granular temperature.

\section{Simulation}
A system consists of two granular particles is used in this work. A particle named wall particle will oscilate with certain frequency $f$ and amplitude $A$. Other particle will be fallen from certain height $h$ and then collide the wall particle and bounce back. The collision and bouncing happen several times and then observed in simulation for a time duration. Both particles have the same mass $m$ and diameter $D$.

A molecular dynamics method (MD) implementing Gear predictor-corrector algorithm \cite{Allen_1999} of 5th order is used in the simulation. Between wall particle and free particle a linear spring-dashpot model is used \cite{Schaefer_1996}.

\begin{figure}[h]
\centering
\includegraphics[width=9cm]{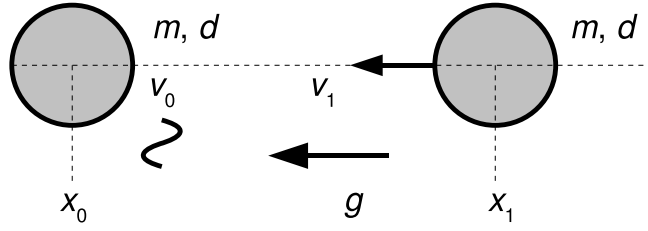}
\caption{\label{fig0} A system consisted of wall particle (left) and free particle (right).}
\end{figure}

Parameters used in the simulation is as followed: $x_0$ = 0, $x_1$ = 4, $m$ = 1, $d$ = 1, $g$ = 0.5, $f$ = 0.1, $v_1(0) = 0$, $k$ = 100, $\gamma$ = 0.1, $\Delta t = 10^{-2}$, $t_i$ = 0, and $t_f$ = 1000. The amplitude $A$ is then varied between 0.03125 and 1.8, which later represented in granular temperature $T_g$ instead of amplitude $A$.

\section{Results and discussion}

As written in \cite{Schaefer_1996}, there is a relation

\begin{equation}
\label{eq1}
\ln \varepsilon = -\frac{\gamma\pi}{\sqrt{4mk - \gamma^2}},
\end{equation}

where is this simulation it it used that $k$ = 100, $\gamma$ = 0.1, and $m$ = 1, so it will be obtained that

\[
\ln \varepsilon = -\frac{0.1}{\sqrt{4 \cdot 1 \cdot 100 - 0.1^2}} = -0.005 \Rightarrow \varepsilon = 0.995,
\]

From \cite{Warr_1995} for sinusoidal wave form

\begin{equation}
\label{eq2}
\frac{4\pi^2Af^2}{g} >> 2.4 (1 - \varepsilon)^{1/2},
\end{equation}

which in this case it is used that $0.03125 \le A \le 1.8$. Left side of Equation (\ref{eq2}) is also known as normalized acceleration \cite{Breu_2003}.

\begin{equation}
\label{eq3}
\Gamma = \frac{4\pi^2Af^2}{g}.
\end{equation}

It is used that $g = 0.5$, it can be calculated that 0.0247 $\le \Gamma \le$ 1.4212. Then right side of Equation (\ref{eq2}) gives 0.170, which tells us that the conducted simulation does not fit the proposed requirement. Even the requirement does not meet it can still be calculated the granular temperature \cite{Warr_1995}.

\begin{equation}
\label{eq4}
T_g = \frac{3.4 m\pi{}Af}{1 - \epsilon}.
\end{equation}

Using Equation(\ref{eq1}) a plot of $x_m$ againts $T_g$ can be obtained, which is in our case

\[
T_g = 213.6A.
\]

\begin{figure}[h]
\centering
\includegraphics[width=10cm]{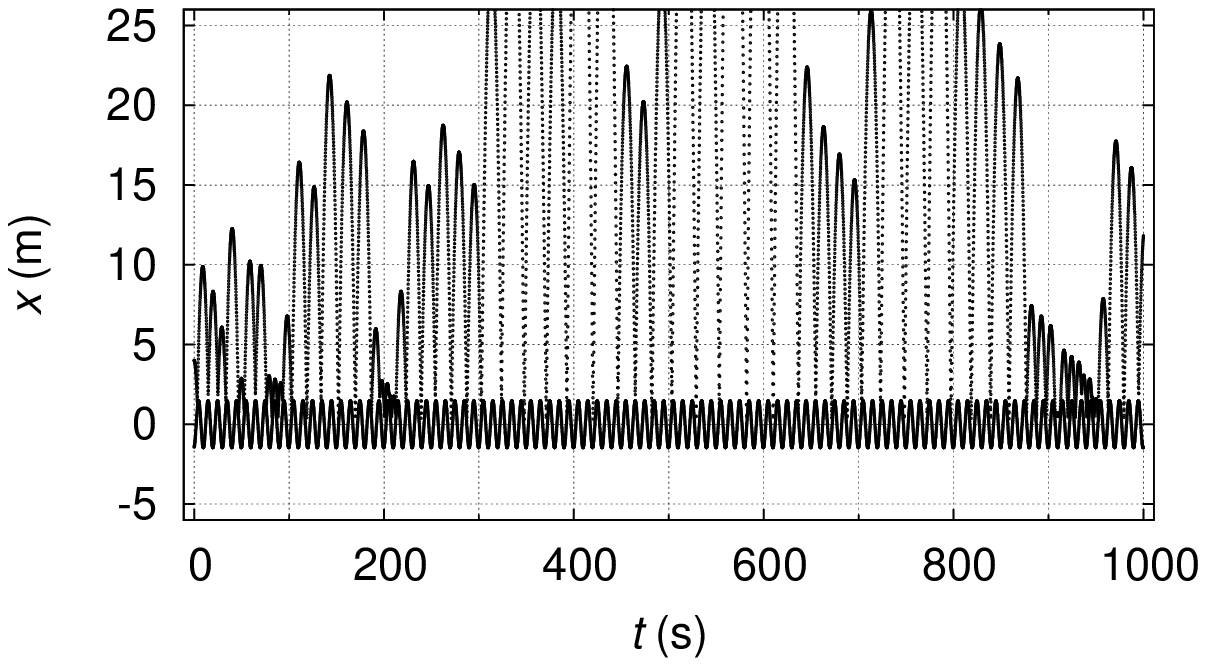}
\includegraphics[width=10cm]{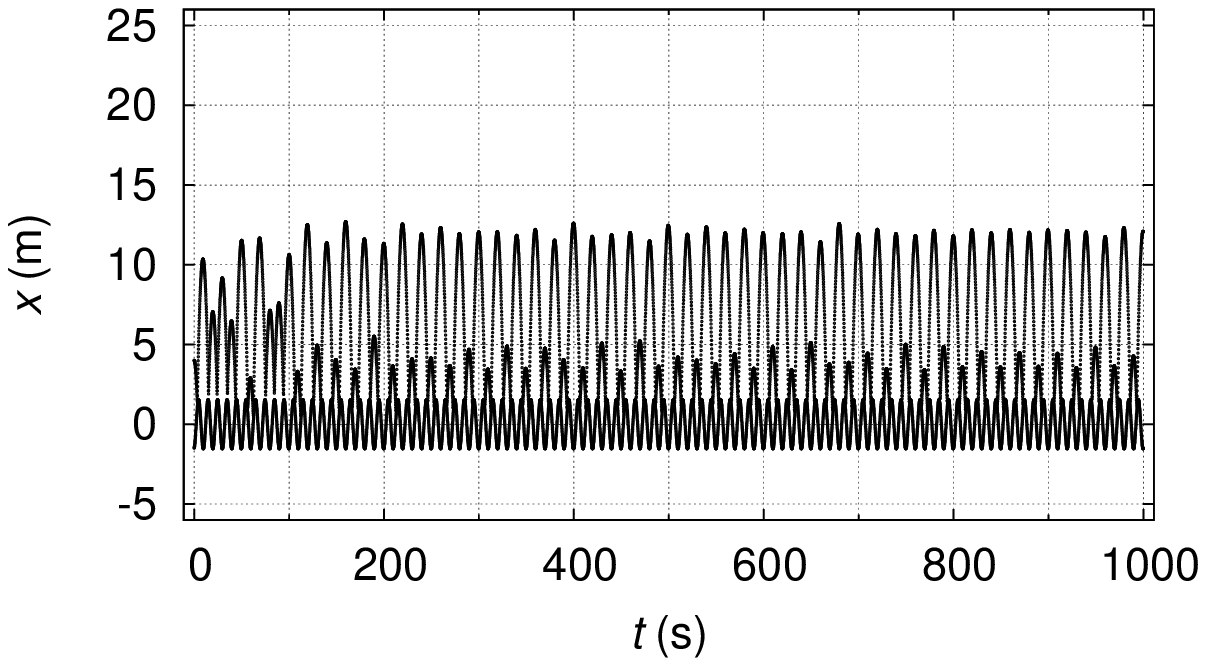}
\caption{\label{fig1} Example of trajectory of bouncing particle at $f = 0.1$ for different amplitude and state: $A$ = 1.43, unsteady (top) and $A$ = 1.50, steady (bottom).}
\end{figure}

\begin{figure}[h]
\centering
\includegraphics[width=10cm]{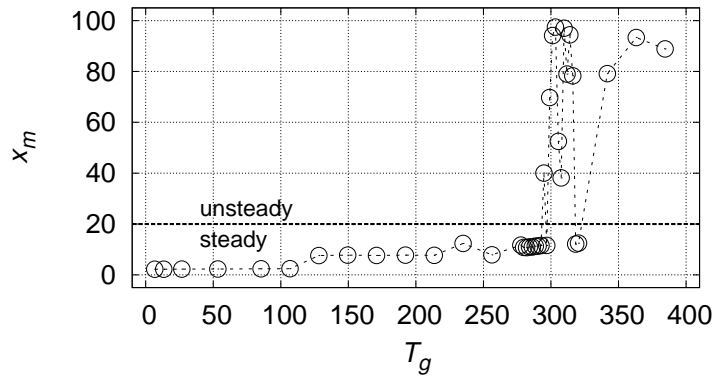}
\caption{\label{fig2} Region of steady and unsteady state (seperated by dashed line) as function of maximum bouncing height $x_m$ and granular temperatur $T_g$.}
\end{figure}

Based on Figure \ref{fig2} it can be said that granular temperature $T_g$ does not play significance role in determining whether the free particle will have a steady or unsteady energy transfer state. These states can be differenced using maximum peak of free particle trajectory $x_m$, which is around 20 as shown in Figure \ref{fig2}.

\section{Conclusions}
Steady and unsteady energy transfer states are defined and can be loosly determined using maximum peak of free particle trajectory. The granular temperature shows no significance in driving the state in which the system stays.

\bigskip\bigskip\noindent
{\bf \Large Acknowledgements}\\ \\
Authors would like to thank to Alumni Association Research Grant in year 2010 for financial support to this work.

\bibliographystyle{unsrt}
\bibliography{manuscript}

\end{document}